\documentclass[prl,aps,twocolumn,showpacs,home,floats]{revtex4}

\usepackage{graphics}
\begin{document}

\title{{\rm\small\hfill (Phys. Rev. Lett., in press)}\\
Kinetic hindrance during the initial oxidation of Pd(100) at ambient pressures}

\author{E. Lundgren}
\email{edvin.lundgren@sljus.lu.se}
\affiliation{Department of Synchrotron Radiation Research,
Institute of Physics, University
of Lund, Box 118, S-221 00 Lund, Sweden}
\author{A. Stierle}
\affiliation{Max-Planck Institut f\"ur Metallforschung,
Heisenbergstr.3, D-70569 Stuttgart, Germany}
\author{M. Todorova}
\affiliation{Fritz-Haber Institut der Max-Planck Gesellschaft,
Faradayweg 4-6, D-14195 Berlin, Germany}
\author{J. Gustafson}
\affiliation{Department of Synchrotron Radiation Research,
Institute of Physics, University of Lund, Box 118, S-221 00 Lund,
Sweden}
\author{A. Mikkelsen}
\affiliation{Department of Synchrotron Radiation Research,
Institute of Physics, University of Lund, Box 118, S-221 00 Lund,
Sweden}
\author{J. Rogal}
\affiliation{Fritz-Haber Institut der Max-Planck Gesellschaft,
Faradayweg 4-6, D-14195 Berlin, Germany}
\author{K. Reuter}
\affiliation{Fritz-Haber Institut der Max-Planck Gesellschaft,
Faradayweg 4-6, D-14195 Berlin, Germany}
\author{J.N. Andersen}
\affiliation{Department of Synchrotron Radiation Research,
Institute of Physics, University of Lund, Box 118, S-221 00 Lund,
Sweden}
\author{H. Dosch}
\affiliation{Max-Planck Institut f\"ur Metallforschung,
Heisenbergstr.3, D-70569 Stuttgart, Germany}
\author{M. Scheffler}
\affiliation{Fritz-Haber Institut der Max-Planck Gesellschaft,
Faradayweg 4-6, D-14195 Berlin, Germany}

\begin{abstract}
The oxidation of the Pd(100) surface at oxygen pressures in the
10${}^{-6}$ to 10${}^3$\,mbar range and temperatures up to 1000 K
has been studied {\em in-situ} by surface x-ray diffraction (SXRD)
The results provide direct structural information on the phases
present in the surface region and on the kinetics of the oxide
formation. Depending on the $(T,p)$ environmental conditions we
either observe a thin $(\sqrt{5} \times \sqrt{5})R27^{\circ}$
surface oxide or the growth of a rough, poorly ordered bulk oxide
film of PdO predominantly with (001) orientation. By either
comparison to the surface phase diagram from first-principles
atomistic thermodynamics or by explicit time-resolved measurements
we identify a strong kinetic hindrance to the bulk oxide formation
even at temperatures as high as 675\,K.
\end{abstract}

\date{\today}



\maketitle

Many technologically important materials containing transition
metals are intended for use under oxygen pressures much higher
than those of the high or ultra high vacuum (UHV) environment
typically employed in surface science related investigations of
the structural, electronic, and chemical properties of these
materials. As the surface properties of such materials may be
significantly altered by the oxidation or corrosion
\cite{over00,hendriksen02,stampfl02} occurring at oxygen pressures
difficult to achieve in conventional UHV experiments, it is
disconcerting that most of our present atomic-scale knowledge
derives from such experiments or from theoretical treatments which
neglect the surrounding gas-phase. Despite the pressure
limitations a few such experiments and theoretical simulations
have in recent years significantly advanced our atomic-scale
understanding of the initial oxidation of metal surfaces,
demonstrating e.g. how radically a surface may change its
structure and functionality under conditions appropriate for high
pressure oxidation catalysis \cite{over00,hendriksen02} and how
the growth of bulk oxide films is often preceded by the formation
of few-atomic-layer-thin so-called surface oxides of complex
geometrical structure and with properties often unknown from the
bulk oxides \cite{carlisle00,lundgren02,todorova03}. The
unexpected and complex behavior revealed by these experiments
emphasizes the need for further {\em in-situ} investigations of
the structural, electronic and chemical surface properties at
higher oxygen pressures but maintaining the accuracy known from
UHV studies - in particular to also address the kinetics of
oxidation and corrosion processes.

A main reason for the lack of {\em in-situ} investigations has
been the scarcity of appropriate experimental techniques that
provide the aspired atomically-resolved information at high
pressure {\em in situ}, or theories that explicitly include the
effect of the surrounding gas-phase. Recently enforced attempts to
overcome this limitation have on the theoretical side led to the
development of first-principles atomistic thermodynamics, where
electronic structure theory calculations are combined with
thermodynamic considerations to address the surface structure and
composition of a metal surface in {\em equilibrium} with arbitrary
environments (See Refs. \cite{reuter02,li03} and references
therein). On the experimental side the state-of-the-art is,
however, currently still characterized by either traditional {\em
ex-situ} atomic-scale investigations \cite{over00}, or new high
pressure techniques like high pressure scanning tunneling
microscopy (STM) that still lack atomic resolution
\cite{hendriksen02}.

In this Letter we demonstrate how precise knowledge of the
$(T,p)$-conditions under which such various oxides, -bulk and
surface-, exist can be obtained by means of {\em in-situ} surface
X-ray diffraction (SXRD) measurements. We have chosen to study Pd
as the metal is a highly active oxidation catalyst of hydrocarbons
under oxygen-rich conditions \cite{ziauddin97}. Still it is
unknown whether Pd or PdO is the active phase \cite{veser99}.
Monitoring the oxidation of the Pd(100) surface over the pressure
range from 10${}^{-6}$ to 10${}^3$\,mbar and up to sample
temperatures of 1000\,K, we observe the formation of both the
previously characterized $(\sqrt{5} \times \sqrt{5})R27^{\circ}$
surface oxide \cite{todorova03} and the transform to
three-dimensional bulk oxide films. Framing our experimental data
with the equilibrium results from atomistic thermodynamics
calculations we can identify kinetically inhibited bulk oxide
growth even at temperatures as high as 675\,K. Under suitable
conditions this hindered transformation to the bulk oxide can be
followed on a time scale currently accessible to the experiment,
opening the door to time-resolved atomic-scale studies of the
initial oxidation process of metal surfaces at ambient pressures.

The SXRD measurements were performed at the Angstr{\o}m Quelle
Karlsruhe (ANKA) beamline in Germany \cite{beamline}. A photon
energy of 10.5\,keV was used and the experiments were conducted in
a six-circle diffraction mode. The crystal basis used to describe
the (H K L) diffraction is a tetragonal basis set \
({\bf{a}$_{1}$},{\bf{a}$_{2}$},{\bf{a}$_{3}$}), with
{\bf{a}$_{1}$} and {\bf{a}$_{2}$} lying in the surface plane and
of length equal to the nearest neighbor surface distance
a$_{\circ}$/$\sqrt{2}$, and a$_{3}$ out-of-plane with length
a$_{\circ}$ (a$_{\circ}$(Pd) = 3.89{\AA}). The UHV x-ray
diffraction chamber allowed partial oxygen pressures of up to
10${}^3$ mbar, and the temperature was estimated from a
thermocouple mounted behind the transferable sample holder,
resulting in an uncertainty of the sample temperature of $\pm
25$\,K. The sample and the cleaning procedure are identical to the
one described in an earlier UHV study \cite{todorova03}. The oxide
films grown were found to be metastable on the time scale of hours
under UHV conditions, and could readily be desorbed at 1175\,K.

The atomistic thermodynamics results are based on
density-functional theory (DFT) calculations performed within the
Full-Potential Linear Augmented Plane Wave (FP-LAPW) scheme
\cite{blaha99} using the generalized gradient approximation (GGA)
\cite{perdew96} for the exchange-correlation functional. The
supercell setup and the highly converged basis-set \cite{basis}
have been detailed in our preceding study \cite{todorova03}. To
determine the range of $(T,p)$-conditions in which this surface
oxide would represent the thermodynamically most stable state, we
evaluate the Gibbs free energy of adsorption \cite{reuter02,li03}
and compare it to other possible states of the system, like the
reported $p(2 \times 2)$ on-surface phase with O in fcc sites
\cite{zheng00,todorova03} or the tetragonal PdO bulk oxide
\cite{rogers71}.

\begin{figure}[t!]
\scalebox{0.28}{\includegraphics{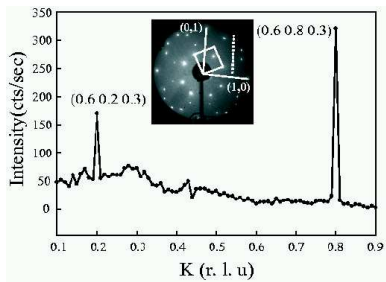}}
\caption{\label{fig1} SXRD in-plane K-scan in reciprocal lattice
units (r.l.u) along (0.6 K 0.3) at $T=575$\,K and
$p=10^{-3}$\,mbar. The location of this scan line in reciprocal
space is indicated in the inset showing the diffraction from the
$(\sqrt{5} \times \sqrt{5})R27^{\circ}$ surface oxide as observed
in UHV-LEED. The direction of the in-plane (H K) basis vectors and
the unit-cell of one of the surface oxide domains are also plotted
in the inset.}
\end{figure}

Figure \ref{fig1} depicts a SXRD K-scan at a sample temperature of
575\,K and partial oxygen pressure of $10^{-3}$\,mbar. From the
inset showing the reciprocal space as observed with UHV low-energy
electron diffraction (LEED) it is obvious that the two observed
peaks in the SXRD scan arise from the $(\sqrt{5} \times
\sqrt{5})R27^{\circ}$ surface oxide. The width of the rocking scan
at the (0.6 0.8 0.3) reflection is (0.13${}^{\circ}$) equal to the
substrate rocking scan in the minimum of the crystal truncation
rod (CTR) before dosing, allowing us to conclude that the surface
oxide domains extend over complete substrate terraces -- in
agreement with our previous STM results \cite{todorova03}.

\begin{figure}[t!]
\scalebox{0.32}{\includegraphics{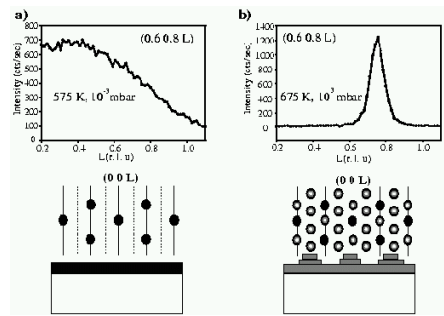}}
\caption{\label{fig2} Top: SXRD out-of-plane L-scan along (0.6 0.8
L). Bottom: corresponding schematic of the observed out-of-plane
diffraction (full lines - rods due to both Pd(100) and the surface
oxide, dashed lines - rods only due to the surface oxide. Bragg
reflections from bulk Pd and bulk PdO are marked as black and grey
circles respectively). a) $p= 10^{-3}$\,mbar and T=575 K, only the
surface oxide diffraction is apparent, b) p= 10${}^3$\,mbar and
T=675 K, diffraction is now due to a bulk-like PdO film.}
\end{figure}

The  out-of-plane diffraction from the (0.6 0.8 L) reflection
under the same temperature and pressure conditions is shown as an
L-scan in the top section of Fig \ref{fig2}a. The smooth decrease
of the intensity and the absence of sharper peaks with increasing
L is a clear fingerprint of a single diffracting layer, in
agreement with the recent finding \cite{todorova03} that the
$(\sqrt{5} \times \sqrt{5})R27^{\circ}$ surface oxide consists of
a single PdO(101) layer adsorbed on the Pd(001) surface. The
observed diffraction changes significantly as the oxygen pressure
and temperature is increased to 10${}^3$\,mbar and 675 K,
respectively, as shown in Fig. \ref{fig2}b. Instead of a smoothly
decreasing diffraction intensity with increasing L, a peak is now
observed at L=0.74. Since the reciprocal lattice is defined by the
lattice constant of Pd, this peak corresponds to a lattice
distance of a$_{\circ}$(Pd)/0.74 = 5.26\,{\AA}, which is very
close to the $c$-lattice constant of bulk PdO, namely 5.33\,{\AA}.
Thus, we observe bulk-like diffraction from PdO, indicating the
formation of several layers of PdO on the Pd(001) surface.
Interestingly, we also note that no diffracted intensity is
observed at L=0 anymore demonstrating that the $(\sqrt{5} \times
\sqrt{5})R27^{\circ}$ surface oxide has completely disappeared
from the surface, i.e. the initially formed PdO(101) plane does
not continue to grow but instead restructures. This is in
agreement with our DFT calculations identifying the (101)
orientation as a higher-energy facet of PdO \cite{todorova03}, and
also with experimental observations on the preferred growth
direction of PdO \cite{mcbride91}. Further, since no finite
thickness oscillations are observed along the rod, the observed
oxide film must be rough, as has also been reported in a recent
high-pressure STM study of this surface \cite{hendriksen03}.

A more detailed analysis of our diffraction data allows us to even
draw some more quantitative conclusions about the properties of
the grown oxide. We observe no change in the overall shape of the
(1 1 L) CTR upon oxidation. Without apparent relaxation, we
therefore attribute the change in integrated intensity at the
minimum (1 1 1) to a change in surface roughness. Estimating the
latter within the $\beta$-model \cite{robinson88} yields
approximately root-mean-square roughness of 10\,{\AA}, whereas the
measured width of $\sim 0.1$ of the L=0.74 peak indicates an
average film thickness of around $\sim 3.89$\,{\AA}/0.1 = $\sim
40$\,{\AA}. The poor order of the formed oxide fringe is further
reflected in the 1${}^{\circ}$ width of the rocking scan at the
L=0.74 peak, indicating either an enhanced mosaic spread, or a
small domain size of the PdO at the surface, which would comply
with the previous STM observations by Hendriksen {\em et al.}
\cite{hendriksen03}. This poor order is probably also reflected by
our inability to detect reflections from the O sublattice. In
fact, the (0.6 0.8 0.74) peak corresponds to the (1 0 1)
reflection in PdO bulk coordinates. In addition to this reflection
we also found the (103), (200) and the (202), and equivalents when
rotating by 90 degrees. By observing that this indexing is similar
to the selection rule for a bcc lattice for which the sum of all
indices must be even, we conclude that what we observe is the Pd
sublattice in PdO, which forms a distorted bcc lattice. The
orientation and geometry of the ordered domains are then
predominantly PdO(001)$\|$Pd(100).

\begin{figure}[t!]
\scalebox{0.32}{\includegraphics{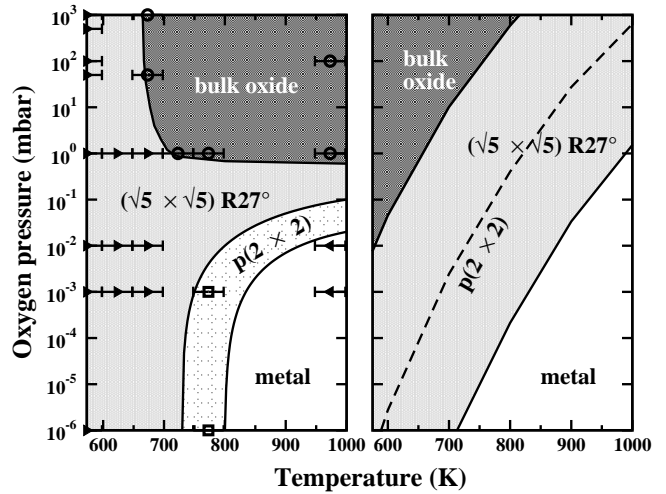}} \caption{\label{fig3}
Left: $(T,p)$-diagram showing the measured phases in the whole
range of experimentally accessible conditions from UHV to ambient
pressure. The ``phase boundaries'' (see text) are rough estimates
to guide the eye. Right: Corresponding surface phase diagram, as
calculated by first-principles atomistic thermodynamics (see
text). The dashed line indicates the thermodynamic stability range
of the $p(2 \times 2)$ adphase, if formation of the surface oxide
was kinetically inhibited.}
\end{figure}

Having thus established the means to distinguish and characterize
the formation of either surface or thicker oxide films at higher O
partial pressures from the diffraction signals, we may in a
straightforward way construct a diagram showing which phase we
measure under which $(T,p)$-conditions. The result is shown in
Fig. \ref{fig3}, covering the whole range of now experimentally
accessible gas pressures from 10${}^{-6}$\,mbar to ambient
pressure. Most strikingly, the $(\sqrt{5} \times
\sqrt{5})R27^{\circ}$ structure is found under a wide variety of
conditions -- even at an oxygen pressure of 10${}^3$\,mbar and a
sample temperature of 575\,K we still observe only the formation
of this surface oxide and no indications for the growth of a
thicker oxide film on the time scale of 1\,hr currently accessible
to the experiment.

As a first step to understand this data we proceed by comparing
it with the surface phase diagram obtained by DFT and atomistic
thermodynamics. Figure \ref{fig3}b gives the corresponding
$(T,p)$-diagram, showing which phase would be most stable on the
basis of our DFT data, if the surface were in full thermodynamic
equilibrium with the surrounding oxygen gas phase. We immediately stress
that although state-of-the-art, some approximations like e.g. the
present treatment of vibrational contributions to the free energies,
as well as the uncertainty introduced by finite basis set and the
employed exchange-correlation functional may well allow for errors
in the phase boundaries of the order of $\pm 100$\,K and (depending on
temperature) of up to several orders in pressure \cite{reuter02,li03}.

Taking this into account we notice a gratifying overall agreement
between theory and experiment in this wide range of environmental
conditions. At a closer look there is, however, a notable
difference that is beyond the uncertainties underlying both the
experimental and theoretical approach: the experimental
observation of the $(\sqrt{5} \times \sqrt{5})R27^{\circ}$ surface
oxide in the top left corner of the drawn diagram, i.e. at lower
temperatures and high pressures. Since the central assumption of
theory, which predicts the stability of the bulk oxide under such
conditions, is the full thermodynamic equilibrium between surface
and gas phase, we interpret this difference as reflecting kinetic
limitations to the growth of the bulk oxide under such conditions.
This is also apparent even within the experimental data set alone.
If the surface was fully equilibrated with the environment, the
evaluated phase boundaries would have to follow lines of constant
oxygen chemical potential, which is the single determining
quantity if thermodynamics applies. In the drawn $(T,p)$-plots
such lines of constant chemical potential would always be parallel
to the phase boundaries as drawn in the theoretical diagram, cf.
Fig. \ref{fig3}b, which the bulk/surface oxide boundary drawn in
Fig. \ref{fig3}a (even considering all uncertainties) is clearly
not. Similarly we also understand the experimental observation of
the $p(2 \times 2)$ adsorbate phase, which according to theory is
never a thermodynamically stable phase, as a sign of kinetic
hindrance to the formation of the $(\sqrt{5} \times
\sqrt{5})R27^{\circ}$ phase. Correspondingly, we have also marked
in the theoretical plot the area, where the $p(2 \times 2)$ would
turn out more stable than the clean surface, if the surface oxide
can not form.

\begin{figure}[t!]
\scalebox{0.29}{\includegraphics{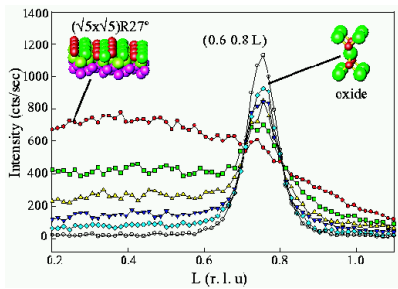}}
\caption{\label{fig4} Six consecutive L-scans along (0.6 0.8 L) at
675\,K and 50\,mbar. Each scan took 240 sec to record, showing the
gradual transformation from the diffraction signal characteristic
of the surface oxide to the one of the bulk oxide, cf. Fig.
\ref{fig2}. Solid circles: first spectrum; empty circles: last
spectrum taken.}
\end{figure}

Under suitable conditions we can even follow the kinetically
limited transition from surface to bulk oxide within the time
resolution of our current experiment. This is illustrated in Fig.
\ref{fig4} showing how the diffraction signal at 675\,K and $p =
50$\,mbar slowly transforms at every consecutive scan from the one
characteristic for the surface oxide phase, cf. Fig. \ref{fig2}a,
to the one of the bulk oxide film, cf. Fig. \ref{fig2}b.
Obviously, this transformation will be faster for higher
temperatures, so that we immediately observe the bulk oxide, or
eventually so slow at lower temperatures that we can no longer
measure the transition within the time frame open to our
experiment. In passing we finally note that these clear kinetic
limitations at time scales of 1\,hr even at technologically
relevant temperatures as high as 675\,K, are somewhat at variance
with the well-known theoretical notion by King and coworkers
suggesting that oxide growth should immediately set in as soon as
it is thermodynamically possible \cite{carlisle00b}.

In conclusion we have studied the oxidation of the Pd(100) surface
from $10^{-6}$\,mbar to ambient pressure by {\em in-situ} SXRD.
Depending on the environmental conditions we observe either the
formation of the $(\sqrt{5} \times \sqrt{5})R27^{\circ}$ surface
oxide (essentially a well-ordered layer of PdO(101)), or the
growth of  $\sim 40$\,{\AA} poorly ordered and rough PdO bulk
oxide, predominantly with PdO(001) orientation. The range of
$(T,p)$-conditions where we measure the surface oxide is
surprisingly large, and goes for $T < 600$\,K even up to ambient
pressures. Comparing with the theoretical surface phase diagram
from first-principles atomistic thermodynamics we interpret this
as reflecting a kinetic hindrance to the formation of the bulk
oxide, which is clearly an activated process due to the involved
massive restructuring at the surface. Such kinetic limitations in
the initial oxide formation process have hitherto barely been
addressed, but could be crucial for understanding high pressure
applications like e.g. oxidation catalysis. Our measurements
demonstrate the usefulness of SXRD for the study of the oxidation
process of almost any material in such high pressure environments,
providing the aspired atomically resolved structural information
of any thin film or nano-based structure exposing its surface to
an ambient working atmosphere.

This work was financially supported by the
Swedish Natural Science Council and the DFG-priority
program ``Realkatalyse''.

\end{document}